\documentclass[preprint,12pt]{elsarticle}

\usepackage[T1]{fontenc}
\usepackage{graphicx}
\usepackage{dcolumn}
\usepackage{bm}

\begin{document}

\title{The Quest for the Heaviest Uranium Isotope}

\author{S. Schramm}
 \ead{schramm@fias.physik.uni-frankfurt.de}
\address{FIAS, CSC, Johann Wolfgang Goethe University,
Frankfurt am Main, Germany}
\author{D. Gridnev}
\address{FIAS, Johann Wolfgang Goethe University,
Frankfurt am Main, Germany}
\author{D. V. Tarasov}
\address{NSC, Kharkov Institute of Physics and Technology, Ukraine}
\author{V. N. Tarasov} 
\address{NSC, Kharkov Institute of Physics and Technology, Ukraine}
\author{W. Greiner}
\address{FIAS, Johann Wolfgang Goethe University,
Frankfurt am Main, Germany}
\date{\today}

\begin{abstract}

We study Uranium isotopes and surrounding elements at very large neutron number excess.
Relativistic mean field and Skyrme-type approaches with different parametrizations are adopted in the study.
Most models show clear indications for isotopes that are stable with respect to neutron emission 
far beyond N=184 up to the range of around N=258.
\end{abstract}

\maketitle

\section{Introduction}
One of the foremost topics in modern nuclear physics is the study of exotic nuclei far beyond the stability line.
A wide range of current experimental activities as well as new upcoming facilities, like FAIR at GSI and FRIB at MSU,
will expand the knowledge of radioactive nuclei up to or close to the neutron drip line.
A key issue of these investigations is to deepen the understanding of the nuclear interactions at large isospin asymmetries, 
which could also yield important input for the study of neutron stars, as the most neutron-rich system known.
The knowledge of the properties, especially binding energies and estimates for lifetimes of neutron-rich isotopes is also
crucial input for r-process nucleosynthesis calculations.
In recent years there have been various studies of the structure of the neutron drip line \cite{Dobaczewski1984,Toki1991,Dobaczewski1994,Hirata:1991,Dobaczewski1995,TERASAKI1996,Nayak,Todd2003,Caurier2001,CAURIER2004,Hilaire2007}. In the latter publication \cite{Hilaire2007} particle-stable Uranium isotopes up to a neutron number of 204 were calculated. 
A number of recent studies suggest a drip  line that might be less smooth as originally expected.
Results show potential peninsulae of isotopes sticking out into the area of unstable nuclei \cite{Gridnev2005,Tarasov2010,Gridnev2006,Tarasov2008}. Other calculations suggest islands of  (meta-)stable nuclei beyond the drip line \cite{Satpathy2006}.
\\ \\
In the following we study very heavy nuclei, from Uranium to Rutherfordium and investigate the final neutron number that can be bound by the nucleus.
The results show nuclei with extremely large neutron excess that are still stable with respect to neutron emission.
This behavior is seen in a number of model approaches and parameter sets.
\section{Theoretical Approach}
The calculational framework used is a relativistic mean-field model (RMF) with a parametrization fitted to binding energies, diffraction and rms radii, and surface thicknesses of a set of nuclei \cite{Bender:1999} as well as non-relativistic Hartree-Fock
Skyrme-type models for various sets of Skyrme forces that were also used in the study of neutron-rich nuclei. 
\\ \\The calculations were performed in spherical symmetry as well as in runs with deformation assuming axial symmetry. The equations were solved on a spatial grid in one and two dimensions to model the nuclei. In comparison the Skyrme calculations were also done in a self-adapting harmonic oscillator basis, where the oscillator asymmetries are adjusted iteratively to minimize the binding energy by allowing for nuclear deformations. 
\\ \\
We employed a BCS-type pairing, where we used delta force pairing for the relativistic and Skyrme calculations on the grid and constant force for the calculations using a harmonic oscillator basis. Certainly, improvements of the pairing description should be done in future investigations, however, the basic qualitative results most likely will not be affected. 
\\ \\
\begin{figure}
\begin{center}
\resizebox{\columnwidth}{!}
{%
\includegraphics{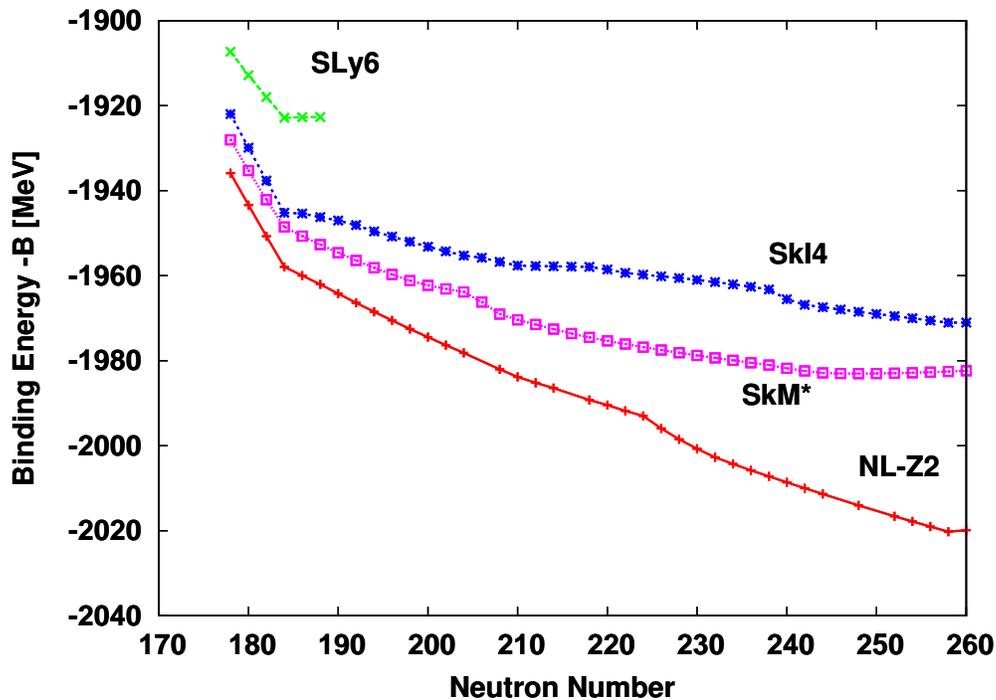}}
\caption{Total binding energy of Uranium isotopes for different parameterizations. The calculation was done for spherical nuclei.
The curves show the result for a relativistic mean-field (NL-Z2) \cite{Bender:1999} and Skyrme models (SkM*, SkI4, SLy6).   
as determined on a spatial grid.
 \label{b1d}
}
\end{center}
\end{figure}
\section{Numerical Results}
Our numerical results of a spherical calculation of Uranium isotopes are shown in Fig. 1. One can see the total energy of the nuclei as function
of neutron number. The curve exhibits an absolute minimum at $^{350}U$ (in the case of the relativistic force), signaling the last neutron-stable isotope.
Several results for RMF and Skyrme calculations as well as for a relativistic effective chiral Lagrangian \cite{deformed} have been obtained most of them exhibiting a minimum at large neutron number. In the chiral Lagrangian the baryonic masses are generated via spontaneous symmetry breaking that generate non-vanishing
vacuum expectation values of the scalar fields. The quality of the model in reproducing knwon nuclear structure data is comparable to other relativistic models
\cite{deformed}.
 As it is based on a flavor-SU(3) description the nucleons also couple weakly to the strange scalar field that corresponds to the strange quark condensate.  This approach does not show a significant increase of bound neutrons with Z. Here a shell closure around $N = 256$ only develops at much higher atomic charge of $Z \ge 116$.
\\ \\ 
In order to check for possible effects from the grid size of the calculation we repeated the calculation for various box sizes from 30 fm to 80 fm
in the case of Uranium without observing significant changes of the binding energy beyond the range of standard numerical uncertainties.
\\ \\
\begin{table}
\begin{center}
\begin{tabular}{llll}
\\
Model&N(1d)&N(2d)&B$_{2d}$ [MeV]\\
\hline \\
NL-Z2 \cite{Bender:1999}&258&258&2021.34\\
SkM* \cite{Brack:1985}&246&220&1987.39\\
SkI4\cite{Reinhard:1995}&258&218&1975.41\\
SLy6\cite{Chabanat:1998}&184&206&1928.92\\
$\chi_M$\cite{deformed}&184&184&1957.56
\end{tabular}
\end{center}
\caption{Heaviest Uranium isotopes for 1d and 2d calculations.
\label{umax}
}
\end{table}
\\ \\
\begin{figure}
\begin{center}
\resizebox{\columnwidth}{!}
{%
\includegraphics{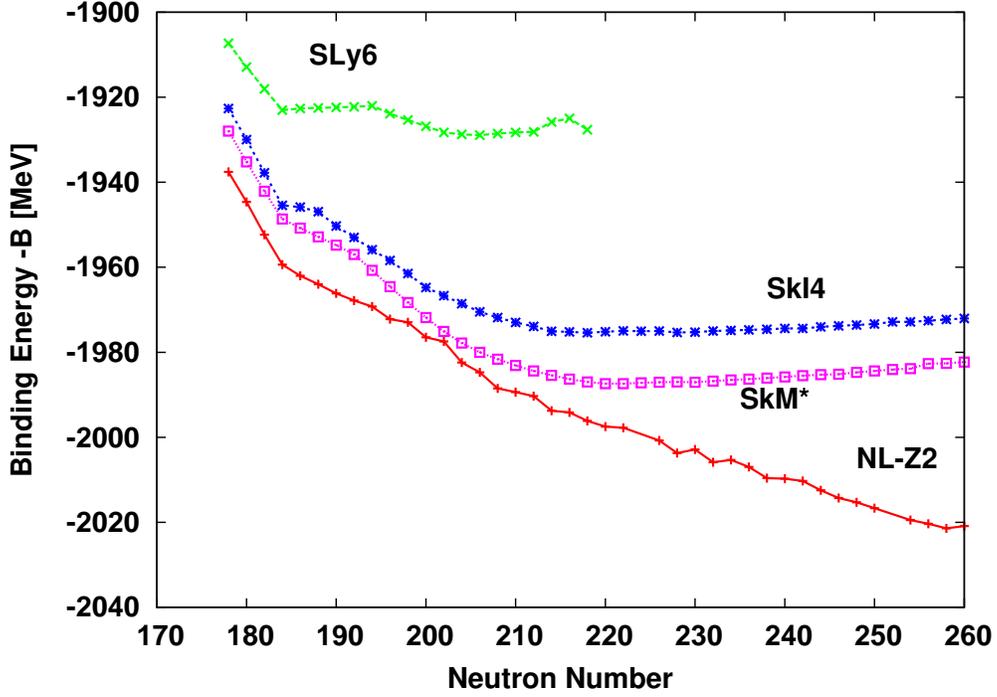}}
\caption{Total binding energy of Uranium isotopes for different parameterizations. The calculation was performed for axially symmetric deformed nuclei
on a two-dimensional spatial grid.
The curves show the result for a relativistic mean-field (NL-Z2) and Skyrme models (SkM*, SkI4, SLy6).   
\label{b2d}
}
\end{center}
\end{figure}
\begin{table}

\begin{center}
\begin{tabular}{lccccc}
\\
Z&NL-Z2&SkM*&SkI4&SLy6&$\chi_M$\\
\hline \\
92&258&220&218&206&184\\
94& 258 & 230 &  230  & 208 &184 \\
96 & 258 & 258 & 230 & 208 &202\\
98 & 258 & 258 & 258 & 218 & 208\\
100 & 258 & 258 & 258 & 220 & 210\\
102 & 258 & 258 & 258 & 220 & 214\\
104 & 258 & 258 & 258 & 232 & 220
\end{tabular}
\end{center}
\caption{Isotopes with maximum binding energy for different models.
\label{nmax}}
\end{table}
\begin{figure}
\begin{center}
\resizebox{\columnwidth}{!}
{%
\includegraphics[angle=-90]{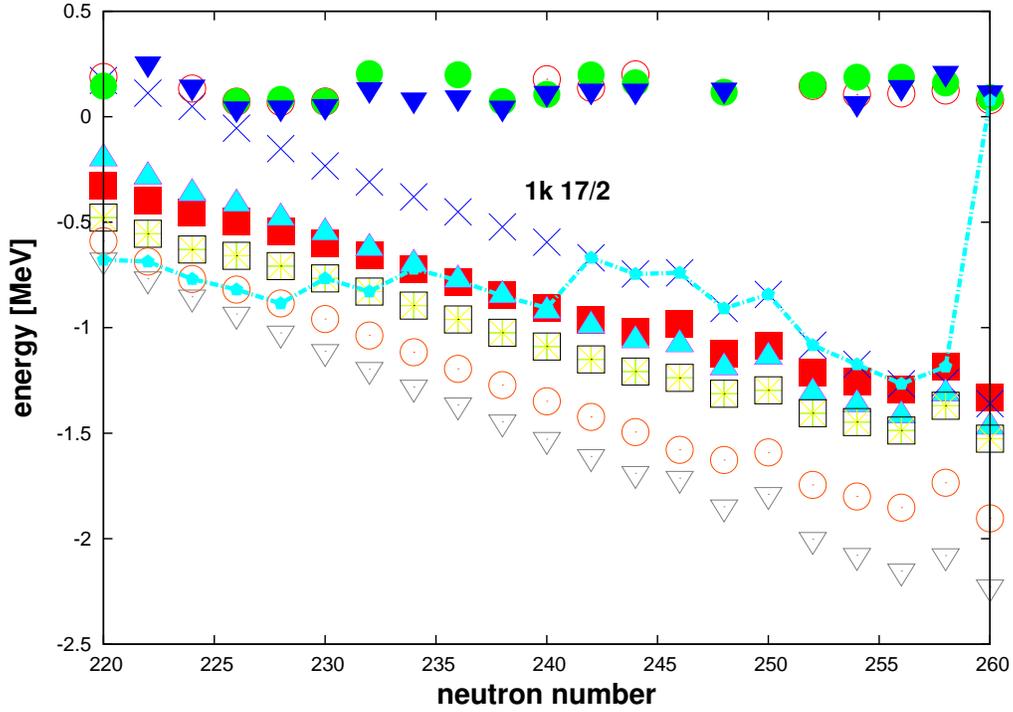}}
\caption{Single-particle levels of neutrons for Uranium isotopes assuming spherical symmetry using the NL-Z2 parametrization.
The line shows the last occupied HF orbit neglecting pairing.  The main cause of the semi-shell closure originates from the filling of the
1k $\frac{17}{2}+$ state. The other bound states shown are (in the order as seen for N=220, starting from the lowest level) 2h $\frac{9}{2}-$, 3f $\frac{7}{2}-$,
4p $\frac{3}{2}-$, 4p $\frac{1}{2}-$, 3f $\frac{5}{2}-$. The 1k $\frac{17}{2}+$ state is only bound for higher values of N as can be seen in the plot. \label{levels}
}
\end{center}:
\end{figure}
\begin{figure}
\begin{center}
\resizebox{\columnwidth}{!}
{
\includegraphics{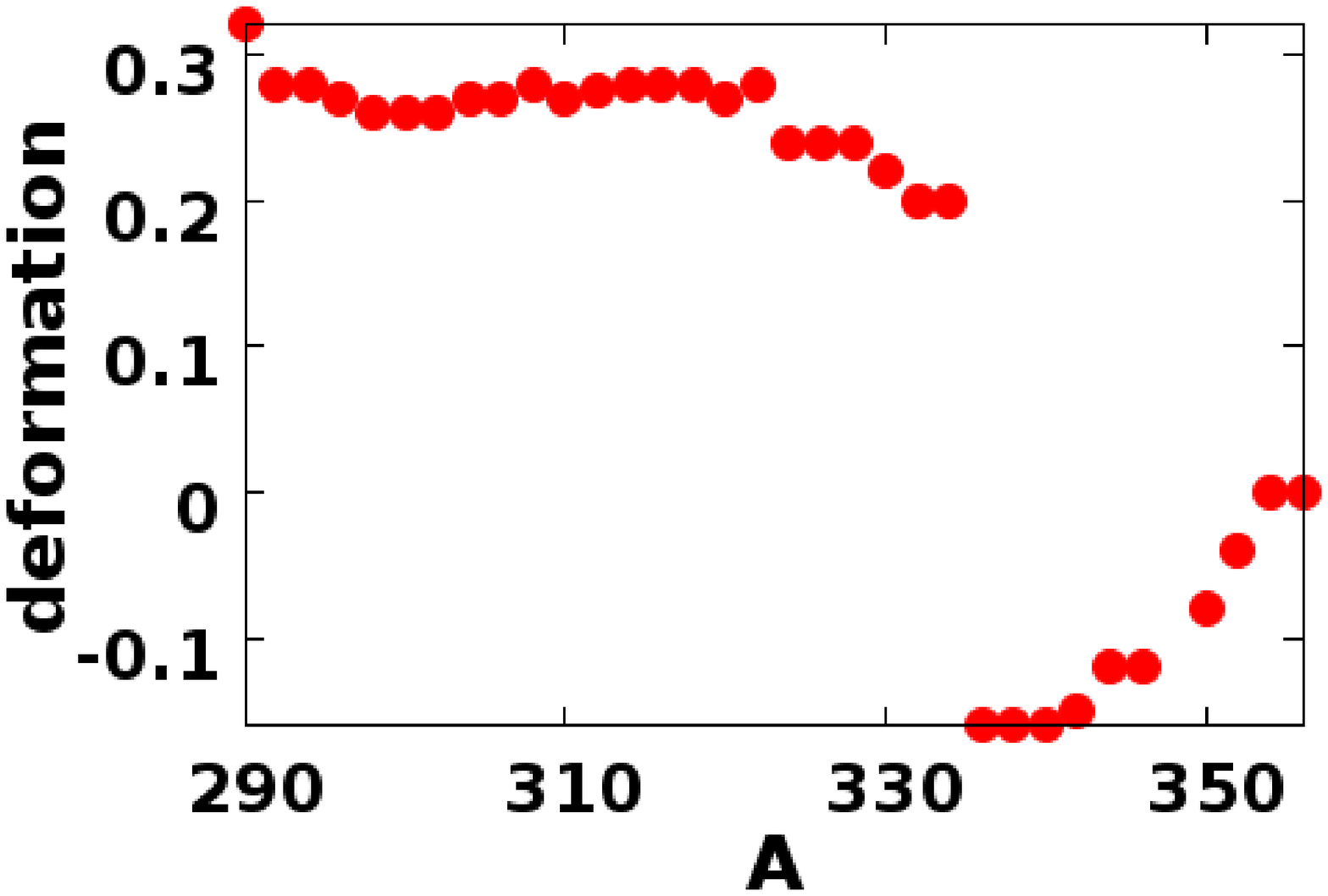}
}
\caption{Deformation $\beta_2$ for different Californium isotopes for the parameter set SkM*.
\label{defCf}
}
\end{center}
\end{figure}
Fig. \ref{b2d} shows the result of a two-dimensional calculation including deformation assuming axial symmetry. The calculations were performed on a spatial grid using three different initial deformations (oblate, spherical, and prolate) for each nucleus. The relativistic force maintains its most bound nucleus for large neutron number N=258, whereas the Skyrme forces show shifted minima as can be observed in the figure.
For comparison, Table \ref{umax} lists the heaviest uranium isotopes for the different parameter sets in spherical approximation and two-dimensional calculations.

The observed behavior is not restricted to  Uranium but hold true for surrounding elements. Table \ref{nmax} shows the most bound isotopes for a range of Z between 92 and 104. Those isotopes exhibit vanishing quadrupole moment. 
\\ \\
The level diagram of the neutron levels from 2.5 MeV below the continuum 
for the case of Uranium isotopes is shown in Fig. \ref{levels} . The $1k17/2^+$ level, which has a high degeneracy, drops with neutron number effectively creating a new shell for large total neutron number of $N = 258$. A similar effect occurs in the case of (some) Skyrme forces as seen in Table \ref{nmax}.
\\ \\
\begin{figure}
\begin{center}
\resizebox{\columnwidth}{!}
{%
\includegraphics[height=2cm]{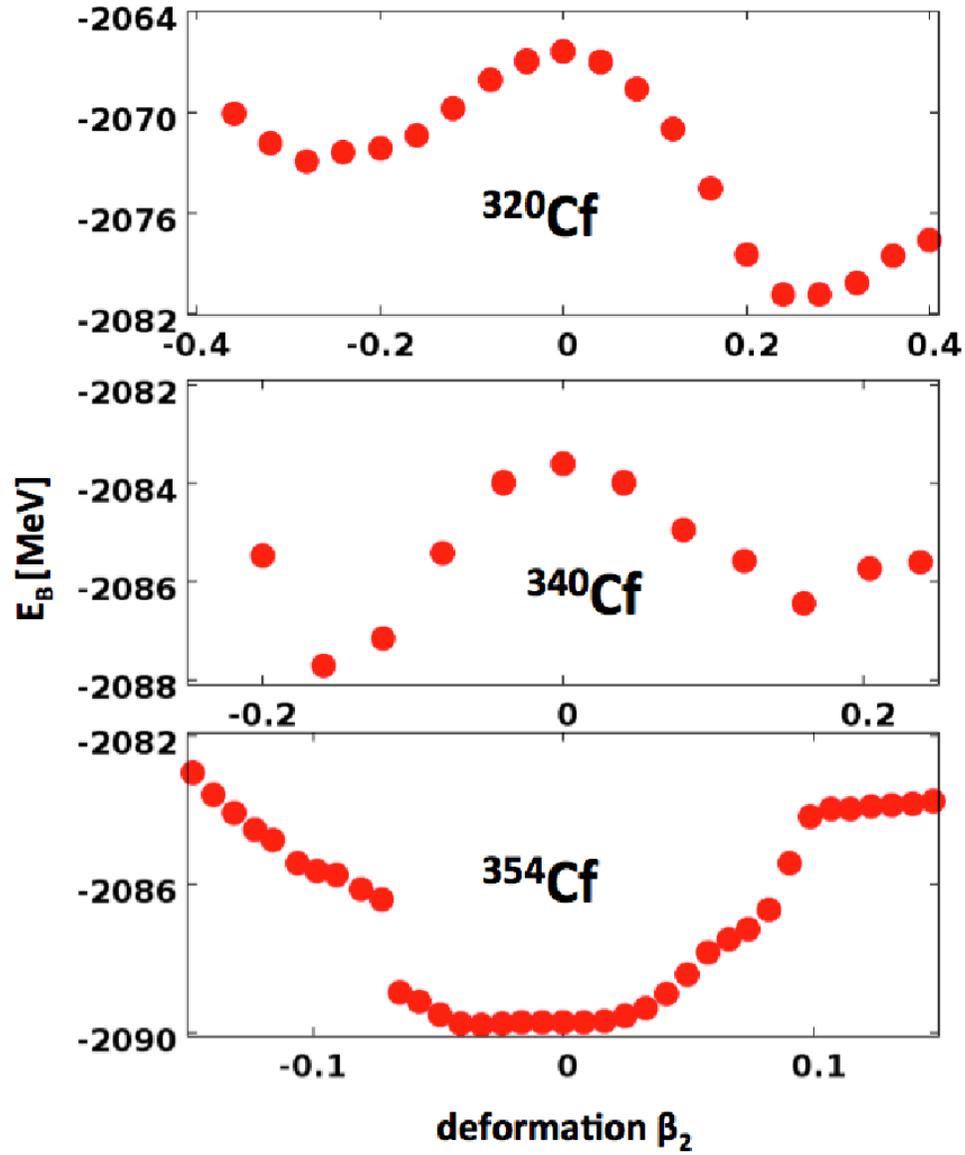}}
\caption{Energy of different Cf isotopes with N = 222, 242, and 256 as function of axial deformation $\beta_2$ .
One can observe the change in deformation from prolate to slightly oblate to spherical. 
\label{Cf}
}
\end{center}
\end{figure}
As an example of the isotope behavior of the nuclear deformation, Fig. \ref{defCf} shows the axial deformation as it changes towards the spherical state at $N=258$ for the parameter set SkM* for Californium isotopes. The binding energy as function of the axial deformation 
for three isotopes along this isotope chain in a calculation including a deformation constraint is depicted in Fig. \ref{Cf}. Here one can observe the the change from prolate to oblate, and 
to finally a spherical state. \\ \\
Looking at the density distribution of protons and neutrons in the case of  the extreme nucleus $^{350}U$ (NL-Z2) one observes an outer shell of essentially 
pure neutron matter as can be seen in Fig.~\ref{rho}.
The neutron skin of this nucleus results as 
\begin{equation}
r_{skin} = r_n - r_p = 7.72\, {\rm fm} - 6.64\, {\rm fm} = 1.08\, {\rm fm}.
\end{equation}
 The isospin asymmetry of such a nucleus
\begin{equation}
 \frac{N-Z}{N+Z} \approx 0.48
 \end{equation}
 is quite substantial compared to a value of 0.7 - 1 inside of a neutron star or a value of less than 0.22 for a $^{208}$Pb nucleus.
\begin{figure}
\begin{center}
\resizebox{\columnwidth}{!}
{%
\includegraphics[angle=0,height=3cm]{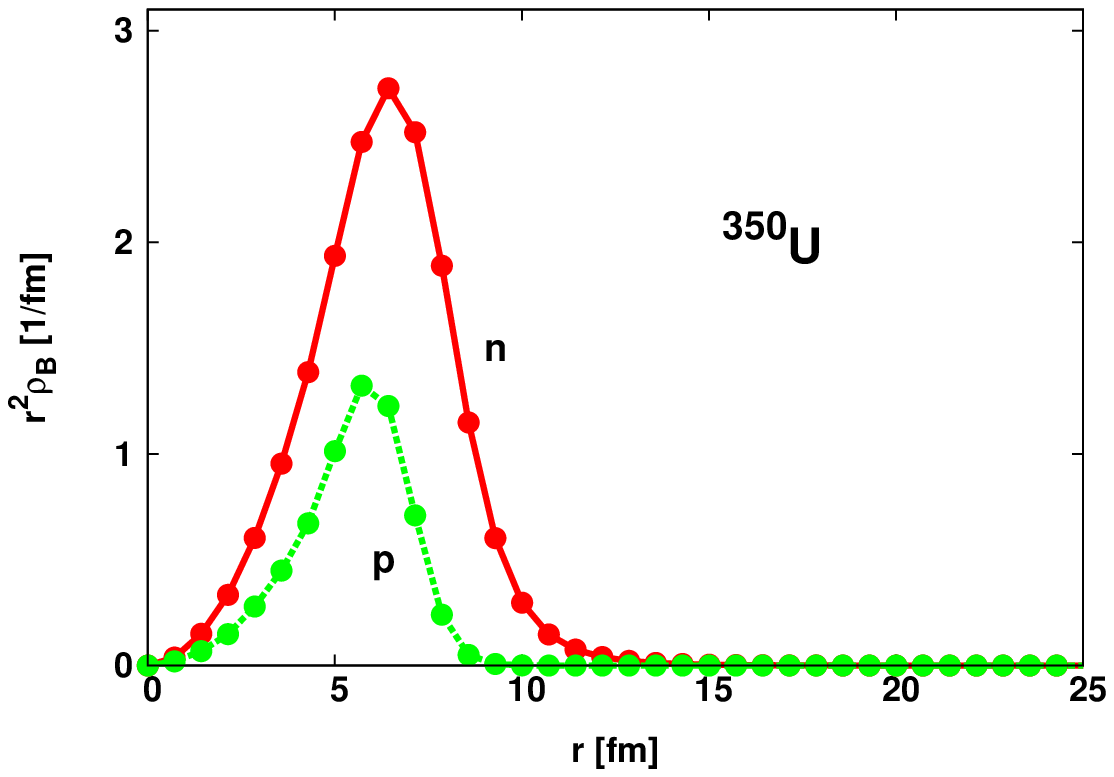}}
\caption{Density distributions $r^2\rho_B$ of protons and neutrons in $^{350}$U for the parameter set NL-Z2.  
\label{rho}
}
\end{center}
\end{figure}
\\ \\
Fig.\ref{Esep} shows the one and two-neutron separation energies $S_n$ and $S_{2n}$ at the magic neutron number of $N = 258$.
Here, $S_n$ was estimated using Koopmans' theorem \cite{Koopmans}. The results for a spherical and a two-dinemsional calculation are shown, which are 
in agreement with each other pointing to the fact that at this neutron number the nuclei are spherical. One can observe that for $Z = 92$ $S_{2n}$ just crosses
the zero line yielding nuclei that are stable with respect to 2 (and 1) neutron emission. However, as can be seen in Table \ref{nmax}, for $Z = 96$ this state becomes the most stable nucleus in the isospin chain for this particular Skyrme parametrization. Thus one can observe the development of similar metastable peninsulas along the neutron drip line as had been observed in \cite{Tarasov2007a}.
\section{Conclusions}
We have calculated heavy nuclei with extreme neutron numbers. Nuclei with up to 258 neutrons are shown to be stable in a range of self-consistent 
model calculations. 
\begin{figure}
\begin{center}
\resizebox{\columnwidth}{!}
{%
\includegraphics[height=4cm]{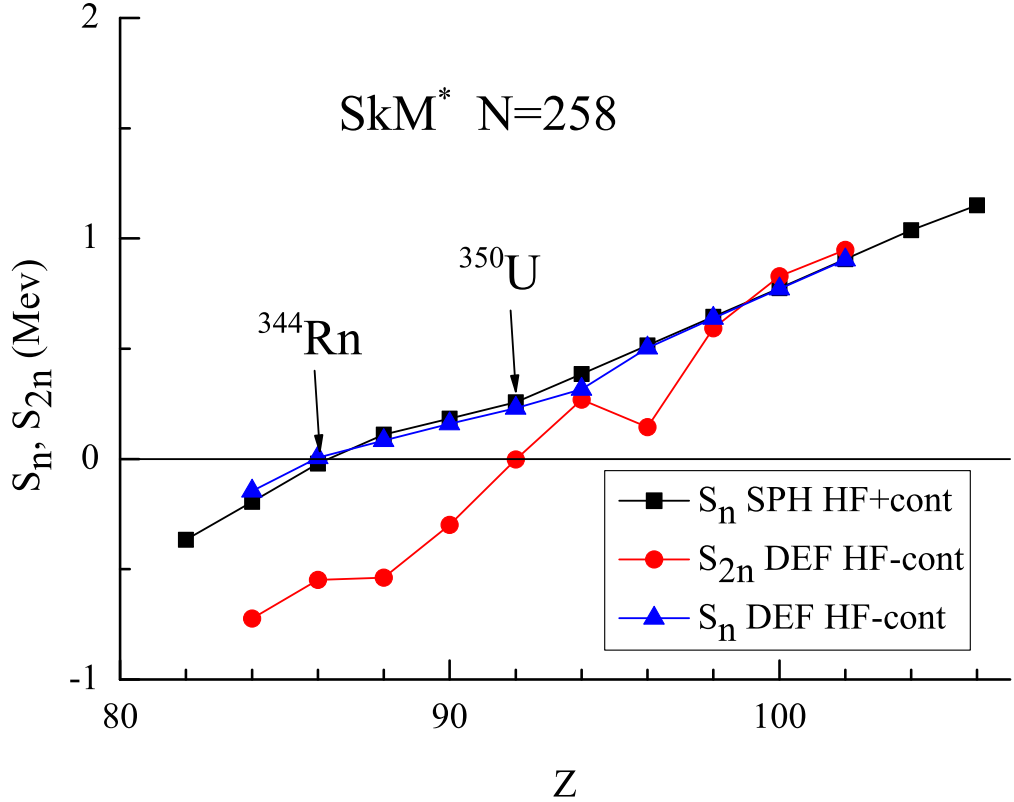}}
\caption{One-neutron ($S_n$) and two-neutron ($S_{2n}$) separation energies  for neutron number N = 184 as function of Z for Skyrme parameters SkM*. The results were 
obtained in a spherical (SPH) and two-dimensional code (DEF) using a basis of oscillator states\label{Esep}.
}
\end{center}
\end{figure}
It is interesting to note that one can achieve self-bound largely neutron system with an asymmetry of 0.48. 
It might be worthwhile to correlate the existence of the states in various models with neutron star observables. At the moment the calculation is largely a theoretical exercise as it is not clear how to produce such nuclei in a realistic experimental setup in measurable quantities, even if, in principle, a lead-lead collision, for example, 
would contain enough neutrons. Recent speculations on using high-intensity pulsed neutron sources from reactors for producing large-N  nuclei \cite{Zag2011} sound like a potentially very interesting new experimental idea for producing very neutron-rich nuclei, but this approach is still very far from practical realization. In the same vein, for now at least there is no scenario predicting production of such states in a realistic astrophysical environment, although similar states might occur in the crust of neutron stars.  However, the knowledge of the boundary of particle-stable nuclei in itself is still a question of fundamental interest in studies of strong-interaction physics.

\bibliographystyle{apsrev4-1}

\begin{thebibliography}{99}
\bibitem{Dobaczewski1984}
J. Dobaczewski, Nucl. Phys. {\bf A422} (1984) 103.
\bibitem{Toki1991}
H. Toki, Nucl. Phys. {\bf A}524 (1991) 633.
\bibitem{Dobaczewski1994}
J. Dobaczewski, I. Hamamoto, W. Nazarewicz, and J. Sheikh, Phys. Rev. Lett. {\bf 72} (1994) 981.
\bibitem{Hirata:1991}
D. Hirata, H. Toki, T. Watabe, I. Tanahita, and B. Carlson, Phys. Rev. {\bf C 44} (1991) 1467. 
\bibitem{Dobaczewski1995}
J. Dobaczewski, W. Nazarewicz, and T. R. Werner, Physica Scripta {\bf T 56}, (1995) 15.
\bibitem{TERASAKI1996}
J. Terasaki, P. Heenen, H. Flocard, and P. Bonche, Nucl. Phys. {\bf A600} (1996) 371.
\bibitem{Nayak}
L. Satpathy and R. C. Nayak, J. Phys. {\bf G 24} (1998) 1527.  
\bibitem{Todd2003}
B. Todd and J. Piekarewicz, Phys. Rev. {\bf C 67} (2003) 044317.
\bibitem{Caurier2001}
E. Caurier, Nucl. Phys. {\bf A693} (2001) 374.
\bibitem{CAURIER2004}
E. Caurier, F. Nowacki, and A. Poves, Nucl. Phys. {\bf A742}  (2004) 14.
\bibitem{Hilaire2007}
S. Hilaire and M. Girod, EPJ {\bf A 33} (2007) 237.
\bibitem{Gridnev2005}
K. A. Gridnev, D. K. Gridnev, V. G. Kartavenko, V. E. Mitroshin, V. N. Tarasov, D. V. Tarasov, and W. Greiner,
EPJ {\bf A 25} (2005) 353.
\bibitem{Tarasov2010}
V. N. Tarasov, K. A. Gridnev, D. K. Gridnev, V. I. Kuprikov, D. V. Tarasov, W. Greiner, and X. Vinyes,
Bull. Russ. Acad. Sci. Phys. {\bf 74} (2010) 1559.
\bibitem{Gridnev2006}
K. A. Gridnev, D. K. Gridnev, V. G. Kartavenko, V. E. Mitroshin, V. N. Tarasov, D. V. Tarasov, and W. Greiner,
Phys. Atom Nucl. {\bf 69} (2006) 1.
\bibitem{Tarasov2008}
V. N. Tarasov, D. V. Tarasov, K. A. Gridnev, D. K. Gridnev, V. G. Kartavenko, and W. Greiner,
IJMPE {\bf 17} (2008) 1273.
\bibitem{Satpathy2006}
L. Satpathy, S. K. Patra, and R. K. Choudhury, arXiv:0604027 (2006) [nucl-th].
\bibitem{Bender:1999}
M. Bender, K. Rutz, P.-G. Reinhard, J. Maruhn, and W. Greiner, 
 Phys. Rev. {\bf C 60} (1999) 034304. 
\bibitem{deformed}
S. Schramm,
Phys. Rev. {\bf C 66} (2002) 064310. 
\bibitem{Brack:1985}
M. Brack, C. Guet, and H. B. Hakansson, 
 Phys. Rep. {\bf 123} (1985) 275. 
\bibitem{Reinhard:1995}
P. Reinhard, 
Nucl. Phys. {\bf A584} (1995) 467.
\bibitem{Chabanat:1998}
E. Chabanat, P. Bonche, P. Haensel, J. Meyer, and R. Schaeffer, 
Nucl. Phys. {\bf A635} (1998) 231.
\bibitem{Koopmans}
T. Koopmans, Physica {\bf 1} (1933) 104.
\bibitem{Tarasov2007a}
V. N. Tarasov, D. V. Tarasov, K. A. Gridnev, D. K. Gridnev, V. G. Kartavenko, W. Greiner, and V. E. Mitroshin
Bull. Russ. Acad. Sci. Phys. {\bf 71} (2007) 747.
\bibitem{Zag2011}
V. Zagrebaev and W. Greiner, 
Phys. Rev. {\bf C 83} (2011) 044618. 
\end{thebibliography}

 \end{document}